\documentclass[twocolumn]{autart}    

\usepackage[english]{babel}
\usepackage{graphicx,epstopdf}         
\usepackage{comment}
\usepackage{dsfont}
\usepackage{bm}
\usepackage{mathrsfs}
\usepackage{amsfonts}
\usepackage{amssymb}
\usepackage{amsmath}
\usepackage{subfigure}
\usepackage{enumerate}
\usepackage[noadjust]{cite}
\usepackage{xcolor}
\usepackage{algpseudocode}
\usepackage{algorithmicx,algorithm}
\usepackage{diagbox}

\definecolor{mblue}{rgb}{0,0.267,0.486}

\newcommand{\Rmnum}[1]{$#1$}
\makeatletter
\newcommand*{\bigs}[1]{{\hbox{$\left#1\vbox to24\p@{}\right.\n@space$}}}
\makeatother

\begin{document}
\begin{frontmatter}

\title{Market Implementation of Multiple-Arrival Multiple-Deadline Differentiated Energy Services \thanksref{footnoteinfo}} 

\thanks[footnoteinfo]{This work was partially supported by the Research Grants Council of Hong Kong Special Administrative Region, China, under the Theme-Based Research Scheme T23-701/14-N.}

\author[HKUST]{Yanfang Mo}\ead{ymoaa@connect.ust.hk},    
\author[PKU]{Wei Chen}\ead{w.chen@pku.edu.cn},               
\author[HKUST]{Li Qiu}\ead{eeqiu@ust.hk},  
\author[UCB]{Pravin Varaiya}\ead{varaiya@berkeley.edu}

\address[HKUST]{Department of Electronic and Computer Engineering, The Hong Kong University of Science and Technology, \\Clear Water Bay, Kowloon, Hong Kong, China}
\address[PKU]{Department of Mechanics and Engineering Science and Beijing Innovation Center for Engineering Science and Advanced Technology, Peking University, Beijing, China, Beijing 100871, China }
\address[UCB]{Department of Electrical Engineering and
Computer Sciences, University of California, Berkeley, CA 94720, USA}


\vspace{-20pt}
\begin{abstract}                          
 An increasing concern in power systems is how to elicit flexibilities in demand, which leads to nontraditional electricity products for accommodating loads of different flexibility levels. We have proposed Multiple-Arrival Multiple-Deadline~(MAMD) differentiated energy services for the flexible loads which require constant power for specified durations. Such loads are indifferent to the actual power delivery time as long as the duration requirements are satisfied between the specified arrival times and deadlines. The focus of this paper is the market implementation of such services. In a forward market, we establish the existence of an efficient competitive equilibrium to verify the economic feasibility, which implies that selfish market participants can attain the maximum social welfare in a distributed manner. We also show the strengths of the MAMD services by simulation. \vspace{-10pt}
\end{abstract}

\end{frontmatter}
\setlength{\belowdisplayskip}{3pt}\setlength{\abovedisplayskip}{3pt}
\section{Introduction}
More and more renewable energy is being absorbed into power systems, which has made the supply/demand balance more difficult. From the supplier's perspective, a large number of reserves should be built to compensate for the volatility of renewable generation. Such an approach is at the expense of economic and environmental benefits~\cite{kirschen2018fundamentals}, so increasing attention has been paid to leveraging flexibilities in demand~\cite{albadi2007demand,callaway2009tapping,siano2014demand}. This is referred to as the~\emph{demand response}, as officially defined by the Federal Energy
Regulatory Commission of the USA, or \emph{demand-side management}, as introduced by the Electric Power Research Institute in the 1980s~\cite{balijepalli2011review}. For example, without compromising on functionality, the charging processes of Electric Vehicles~(EVs) and residential pool pumps~\cite{gan2013optimal,d2015demand,ma2016efficient}, can be modulated, suspended, and/or resumed to match supplies. It was shown in~\cite{meyn2015ancillary} that flexible loads can provide ancillary regulation in a proper frequency band to maintain power system stability. More relevant results can be found in~\cite{do2018stochastic,mo2016duration,singh2018decentralized,zhu2016evolutionary}. 

Apart from technical matters, we are interested in economic issues; e.g., what electricity products are capable of eliciting load flexibilities? Traditionally, we mostly treat electrical energy as a homogeneous product sold at a unit price, while in the demand response, we can classify electricity services into distinct energy products according to their different levels of flexibility, which are often referred to as differentiated electricity services~\cite{oren2012service}. See, for instance, the products in~\cite{tan1993interruptible} and~\cite{bitar2017deadline}.

Along this line, we introduce Multiple-Arrival Multiple-Deadline~(MAMD) differentiated energy services in our conference paper~\cite{mo2017differentiated}. Such services are designed for the flexible loads that require constant power for specified durations but are indifferent to the actual power delivery time as long as the duration requirements are satisfied between the specified arrival times and deadlines. In this case, a load is more flexible if it requires a shorter duration, earlier arrival time, or later deadline.

If all the loads share the same arrival time, then the MAMD model reduces to the duration-deadline jointly differentiated energy services studied by the authors in~\cite{chen2015duration} and~\cite{chen2016constrained}. If we further require all the loads to have the same deadline, then the MAMD model reduces to the duration-differentiated energy services studied in~\cite{nayyar2016duration} and~\cite{negrete2016rate}. Thanks to these pioneering results, we inquire into the market implementation of MAMD services.

In this paper, we discuss a forward market implementation of MAMD services via two economic issues. One is the social welfare maximization problem, where all the market participants are altruistic and cooperative. The other is the competitive equilibrium, where each member participates rationally in its own interests. We analyze the optimal social welfare obtained by a social planner who makes decisions on behalf of both the supplier and consumers. Furthermore, we prove that the mechanism of this market is in itself capable of leading self-interested consumers to such optimal social welfare. Thus, we verify theoretically the economic feasibility of MAMD services.

The rest of the paper is as follows. In Section~\Rmnum{2}, we describe the MAMD differentiated energy services and revisit the supply/demand matching problem which has been studied in the conference paper~\cite{mo2017differentiated}. We deal with the forward market implementation in Section~\Rmnum{3}, where we successively study the social welfare optimization problem and the competitive equilibrium. After introducing simulation results in Section~\Rmnum{4}, we end this paper with conclusions and future work.

\textsl{Notation}\\
Let~$\mathbb{R}$,~$\mathbb{Z}$, or~$\mathbb{N}$ denote the set of real numbers, integers, or nonnegative integers. For~$n\in \mathbb{N}$, let~$\underline{n}$ denote the set~$\{1,2,\ldots,n\}$. For two sets~$\mathcal{X},\mathcal{Y}$, let~$\mathcal{X}\subseteq \mathcal{Y}$ denote that~$\mathcal{X}$ is a subset of~$\mathcal{Y}$. The cardinality of a set~$\mathcal{X}$ is denoted by~$|\mathcal{X}|$. We use~$O$ and~$E$ to denote a matrix or tensor of a compatible dimension whose elements are all zeros and all ones respectively. For a vector~$\bm x\in \mathbb{R}^n$, we define~$\|\bm x\|_1=\sum_{i=1}^n |x_i|$. For an assertion~$\mathfrak{A}$, the indicator function $\mathds{1}(\mathfrak{A})$ is one if~$\mathfrak{A}$ is true and zero otherwise. Define $[a]^+=\textrm{max}\{a,0\}$ for~$a\in \mathbb{R}$.

\section{MAMD Differentiated Energy Services}
In this section, we elaborate on the MAMD differentiated energy services and the supply/demand matching.

\subsection{Supply/Demand Model Formulation}
{We herein consider the loads which demand a uniform constant power level for specified durations within an operational horizon, which is evenly divided into~$n\in \mathbb{N}$ time slots. For instance, we may charge a battery EV for two hours within a day~($n=24$~hours) at the rate of one unit per hour. Thus, most quantities to be introduced are integers. To avoid complicating the study by details, we shall theoretically analyze an abstract model while leave out practical constraints by referring the readers to~\cite{negrete2016rate} for the techniques of fitting the model to real data.}

After the long-term evolution and observations, the system operator suggests a set of possible arrival times or deadlines, namely,~$\mathcal{T}=\{n_j\in \mathbb{N}\mid 0\leq j \leq \nu \}$, for a certain~$\nu \in \underline{n}$, wherein~$n_0=0$, $n_\nu=n$,~and~\mbox{$n_i<n_j$} whenever~$i<j$. Without loss of generality, we assume that each consumer orders a portion of MAMD differentiated energy service defined by three integer parameters~$\left(r,{a},{d}\right)$, which means that the consumer demands~$1$ unit/slot of power for~$r$ time slots between the~$(n_{a}+1)$th and the~$n_{d}$th time slot. The first parameter~$r$ specifies the charging duration, while the last two,~$a$ and~$d$, respectively specify the arrival time and deadline of the possible power delivery. The flexibility of such a load lies in that it is indifferent to the actual power delivery time, so the power provided by the service~$(r,a, d)$ will be possibly delivered in any~$r$ time slots from the~$(n_{a}+1)$th time slot to the~$n_{d}$th time slot. Given~$\mathcal{T}$, we denote all the MAMD differentiated energy services by~$$\mathcal{S} =\left\{\left(r,{a},{d}\right)\mid r,a,d\in\mathbb{N},  a< d\leq \nu, 0<r \leq {n_d-n_a}\right\}.$$

\begin{figure}[b]
  \begin{minipage}[t]{0.46\linewidth}
    \centering
    \mbox{ \begin{tabular}{c|c}
        \diagbox[height=1.54em,width=4.5em]{\hspace{-2pt}\vspace{-4pt}\small Loads }{\small $\bm h$}  &    $\begin{bmatrix}2&4&2&5&1&3\end{bmatrix}$ \\ \hline
        $\begin{matrix}
          (2,0,2)\\(3,0,2)\\(5,0,3)\\(2,1,3)\\(2,1,2)
        \end{matrix}$ & $\begin{bmatrix}
                          ? & ? & ? & ? & \underline{0} & \underline{0} \\
                          ? & ? & ? & ? & \underline{0} & \underline{0} \\
                          ? & ? & ? & ? & ? & ? \\
                          \underline{0} & ? & ? & ? & ? & ? \\
                          \underline{0} & ? & ? & ? & \underline{0} & \underline{0}
                        \end{bmatrix}$
     \end{tabular} }
  \end{minipage}   ~\vline%
  \begin{minipage}[h]{0.46\linewidth}
    \centering
  \mbox{ \begin{tabular}{c|c}
        \diagbox[height=1.54em,width=4.5em]{\hspace{-2pt}\vspace{-4pt}\small Loads }{\small $\bm h$}  &    $\begin{bmatrix}2&4&2&5&1&3\end{bmatrix}$ \\ \hline
        $\begin{matrix}
          (2,0,2)\\(3,0,2)\\(5,0,3)\\(2,1,3)\\(2,1,2)
        \end{matrix}$ & $\begin{bmatrix}
                          0 & 1 & 0 & 1 & \underline{0} & \underline{0} \\
                          0 & 1 & 1 & 1 & \underline{0} & \underline{0} \\
                          1 & 1 & 0 & 1 & 1 & 1 \\
                          \underline{0} & 0 & 0 & 1 & 0 & 1 \\
                          \underline{0} & 1 & 0 & 1 & \underline{0} & \underline{0}
                        \end{bmatrix}$
     \end{tabular} }
  \end{minipage}
  \caption{Consider~$\mathcal{T}=\{n_0=0,n_1=1,n_2=4,n_3=6\}$. A feasible power allocation matrix is given on the right.}
  \label{fig: flexiblefeasible}
\end{figure}

\vspace{-4pt}
\subsection{Supply/Demand Matching Revisited}
In general, the supply is not unlimited and the operator has to use the limited supply to match a given demand. We define the \emph{supply profile} as~\mbox{$\bm{h}=[h_1~h_2~\cdots~h_n]\in \mathbb{N}^n$}, which means that there are~$h_j$ units of electrical energy available at the~$j$th time slot, for all~$j\in \underline{n}$. We consider~$m$ flexible loads and the~$i$th load claims a portion of the MAMD service~$(r_i,a_i,d_i)\in \mathcal{S}$, for all~$i\in \underline{m}$. To be concise, all the charging durations constitute the~\emph{duration profile}~$\bm{r}=[r_1~r_2~\cdots~r_m].$ Similarly, the arrival times and deadlines of the~$m$ loads are summarized as~${\bm a}$ and~${\bm d}$. In short, we use~$\left(\bm r, {\bm a}, {\bm d}\right)$ to represent the demand of the collection of loads.

We say the supply~$\bm h$ is adequate for the demand~$\left(\bm r, {\bm a}, {\bm d}\right)$ if there exists a feasible power allocation. An allocation is denoted by an~$m \times n$ $(0,1)$-matrix~$A$, where~$A(i,j)$ is one if the~$i$th load will be charged at the~$j$th time slot and zero otherwise. As exemplified in Figure~\ref{fig: flexiblefeasible}, an allocation matrix is feasible if its~$i$th row is consistent with the MAMD service~$(r_i,a_i,d_i)$~(i.e.,~$\|A(i,:)\|_1=r_i$ and~$A(i,j)=0$ if~$j\notin [n_{a_i}+1,n_{d_i}]$) and its~$j$th column sum is no more than the supply~\mbox{$h_j$~(i.e.,~$\|A(:,j)\|_1\leq h_j$}), for all~$i\in \underline{m}$ and~$j\in \underline{n}$. For simplicity, we temporarily ignore the issues from~\mbox{$\|\bm h\|_1>\|\bm r\|_1$} in this paper. In practice, we could sell the redundant supplies to outer grids or handle them by reserves or curtailments of generation. As illustrated in Figure~\ref{fig: flexiblefeasible}, the supply/demand matching is mathematically equivalent to a matrix completion problem concerning the existence of a $(0,1)$-matrix with given row sums, upper-bounded column sums, and predetermined zeros, or in other words, the nonemptiness of the class of all feasible power allocation matrices, denoted by~$\mathcal{A}(\bm h,\bm r, {\bm a},{\bm d})$.

Without loss of generality, we assume the monotonicity throughout this paper: for all~$\kappa\in \underline{\nu}$, we have~$h_i \geq h_j$ whenever~$n_{\kappa-1}+1\leq i < j \leq n_\kappa$.

When~$\nu=1$, the MAMD model reduces to duration-differentiated energy services, i.e.,~${a_i}\!=\!0$ and~${d_i}\!=\!1$, for all~$i\in \underline{m}$. In this case, we use~$\mathcal{A}(\bm h,\bm r, {\bm 0},{\bm 1})$ to denote the feasible power allocation matrix class. The Gale-Ryser theorem~\cite{gale1957theorem,ryser1957combinatorial} states that~$\mathcal{A}(\bm h,\bm r, {\bm 0},{\bm 1})$ is nonempty if and only if the following inequalities hold:~\begin{equation}\label{OneOrderTensor}
  W_k(\bm h,\bm r, {\bm 0},{\bm 1})=\sum_{j=k+1}^{n}h_j-\sum_{i=1}^{m}[r_i-k]^+\geq 0,\end{equation}
where~$0\leq k \leq  n-1$. If we set~$W_n(\bm h,\bm r, {\bm 0},{\bm 1})=0$, then we can check the above inequalities in a recursive way since we show that, for~$k=n,n-1,\ldots,1$ in order,~$$W_{k-1}(\bm h,\bm r, {\bm 0},{\bm 1})=W_k(\bm h,\bm r, {\bm 0},{\bm 1})+h_k-\sum_{i=1}^{m}\mathds{1}\!\left(r_i\geq k\right).$$
For the case where~$\nu \geq 1$, we generalize the left part of~(\ref{OneOrderTensor}) by defining a~$\nu$th-order structure tensor as
\begin{align*}
\begin{split}
&W_{k_1k_2\cdots k_{\nu}}(\bm{h},\bm{r}, {\bm a},{\bm d})=\\
  &\sum_{\kappa=1}^{\nu}\sum_{j=n_{\kappa-1}+k_\kappa+1}^{n_\kappa}\!\!\!\!\!h_j-\!\sum_{i=1}^{m}\left[r_i-k_{a_i+1}-k_{a_i+2}-\cdots-k_{d_i}\right]^+\!, \end{split}
\end{align*}where~$0\leq k_\kappa \leq n_\kappa-n_{\kappa-1}$, for each~$\kappa\in \underline{\nu}$. For notational convenience, we herein let every index~$k_\kappa$ start from~$0$, for all~$\kappa\in \underline{\nu}$. We call it a structure tensor according to tradition~\cite{ryser1960traces,chen2016constrained}. Similarly to before, the tensor can be calculated recursively. Specifically,~$W_{k_1k_2\cdots k_{\nu}}(\bm{h},\bm{r}, {\bm a},{\bm d})=0$, when~$k_\kappa = n_\kappa-n_{\kappa-1}$, for all~$\kappa\in \underline{\nu}$ and
\begin{align*}&W_{k_1\cdots k_{j-1}(k_j-1)k_{j+1} \cdots k_v}(\bm{h},\bm{r}, {\bm a},{\bm d})\\&=W_{k_1\cdots k_j \cdots k_v}(\bm{h},\bm{r}, {\bm a},{\bm d})+h_{n_{j-1}+k_j}\\&-\sum_{i=1}^{m}\mathds{1}\!\left(a_i< j \leq d_i,r_i\geq (k_{a_i+1}+k_{a_i+2}+\cdots+k_{d_i})\right),\end{align*}
for all~$1\leq k_j\leq n_j-n_{j-1}$ and each~$j\in \underline{v}$. As a result, the time complexity of computing~$W(\bm{h},\bm{r}, {\bm a},{\bm d})$ is~$\mathcal{O}\left(m(n_1-n_0+1)\times \cdots \times (n_\nu-n_{\nu-1}+1)\right)$. If every element in the tensor~$W(\bm{h},\bm{r}, {\bm a},{\bm d})$ is nonnegative, then we write~$W(\bm{h},\bm{r}, {\bm a},{\bm d})\geq O$. The following result, firstly recorded in~\cite{mo2017differentiated} without proofs, specifies the condition under which the supply can satisfy the demand via the nonnegativity of the structure tensor.
\begin{thm}\label{tensorthm}
  A supply profile $\bm h$ is adequate for the demand~$\left(\bm r, {\bm a}, {\bm d}\right)$ if and only if~$W(\bm h,\bm r,{\bm a},{\bm d})\geq O$.
\end{thm}\vspace{-15pt}
\begin{pf}
\setlength{\belowdisplayskip}{3pt}\setlength{\abovedisplayskip}{2pt}
\begin{figure}[b]
  \begin{minipage}[t]{0.26\linewidth}
    \centering
   \mbox{ $\begin{matrix}
       (r_1,0,2)\\
       (r_2,0,2)\\
       (r_3,0,3)\\
       (r_4,1,3)\\
       (r_5,1,3)
     \end{matrix}$}
  \end{minipage}  ~\vline~%
  \begin{minipage}[h]{0.69\linewidth}
    \centering
    \includegraphics[scale=0.6]{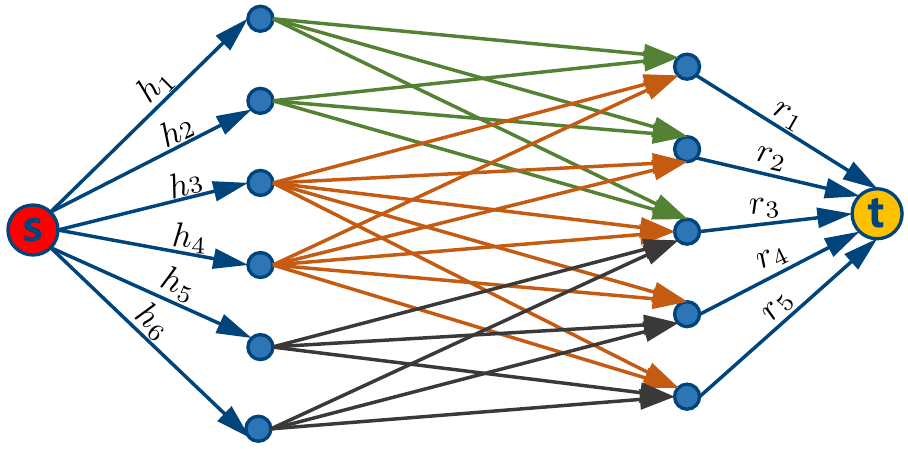}
  \end{minipage}
  \caption{Five loads with~$\mathcal{T}=\{n_0=0,n_1=2,n_2=4,n_3=6\}$, and an associated $s$-$t$ flow network.}
  \label{fig: flowgraph}
\end{figure}
Firstly, we shall construct an associated \mbox{$s$-$t$} network, as depicted in Figure~\ref{fig: flowgraph}. We associate each row/column with an intermediate vertex. For all~$i\in \underline{m}$, we add an arc from the $i$th row vertex to the sink node~$t$ and define its capacity as~$r_i$. For all~$j\in \underline{n}$, we add an arc from the source node~$s$ to the~$j$th column vertex and define its capacity as~$h_j$. In addition, for all~$i\in \underline{m}$ and~$j\in[n_{a_i}+1,n_{d_i}]$, we add an arc from the~$j$th column vertex to the~$i$th row vertex and define its capacity as one. By the Integral Flow theorem~\cite{dantzig1955max}, we observe that~$\mathcal{A}(\bm h,\bm r, {\bm a},{\bm d})$ is nonempty if and only if the associated~$s$-$t$ network has a maximal flow of value~$\|\bm r\|_1$.

Then, by the Max-Flow-Min-Cut theorem~\cite{ford1956maximal}, we conclude that such a flow exists if and only if every \mbox{$s$-$t$} cut has a capacity no less than~$\|\bm r\|_1$ in the associated flow network. There are an exponential number~$(2^{m+n})$ of such $s$-$t$ cuts. We characterize each cut by a subset~$\mathcal{X}$ of~$\underline{m}$ and a subset~$\mathcal{Y}$ of~$\underline{n}$, and denote its capacity by~$c(\mathcal{X},\mathcal{Y})$. Specifically, the capacity~$c(\mathcal{X},\mathcal{Y})$ comes from three elements: the arcs from the source node~$s$ to~the column vertices which are not indexed by~$\mathcal{Y}$, the arcs from the column vertices indexed by~$\mathcal{Y}$ to the row vertices indexed by~$\mathcal{X}$, and the arcs from the row vertices which are not indexed by~$\mathcal{X}$ to the sink node~$t$. Hence, the value of a maximal flow is no less than~$\|\bm r\|_1$ if and only if~$c(\mathcal{X},\mathcal{Y})\geq \|\bm r\|_1$, for every~$\mathcal{X}\subseteq \underline{m}$ and~$\mathcal{Y}\subseteq\underline{n}$, where~\begin{equation}\label{formulacxy}
  \begin{split}
    c(\mathcal{X},\mathcal{Y})=& \sum_{j=1}^nh_j-\sum_{j\in \mathcal{Y}}h_{j}+\sum_{i=1}^mr_i-\sum_{i\in \mathcal{X}}r_{i}\\
 &+\sum_{i\in \mathcal{X}}\sum_{j\in \mathcal{Y}}\mathds{1}(n_{a_i}<j\leq n_{d_i}).
  \end{split}
\end{equation}In what follows, we will remove a number of redundant inequalities above. We observe that for~$\bm x,\bm y\in \mathbb{R}^n$, it holds that~$\max_{\mathcal{Z}\in \underline{n}}\sum_{i\in \mathcal{Z}}(x_i-y_i)=\sum_{i=1}^{n}[x_i-y_i]^+$. Based on this observation, we show that
\begin{multline*}
  w(\mathcal{Y})=\min_{\mathcal{X}\in \underline{m}}c(\mathcal{X},\mathcal{Y})=\sum_{j=1}^nh_j-\sum_{j\in \mathcal{Y}}h_{j}+\sum_{i=1}^mr_i\\
  \ \ \ \ -\sum_{i=1}^m\Bigg[r_{i}-\sum_{j\in \mathcal{Y}}\mathds{1}(n_{a_i}<j\leq n_{d_i})\Bigg]^+.
\end{multline*} Moreover, if we fix the cardinality of~$\mathcal{Y}$ as~$\tau$ ($0\leq \tau \leq n$), then by the monotonicity assumption on~$\bm h$, we obtain
\begin{multline*}
  \min_{|\mathcal{Y}|=\tau}w(\mathcal{Y}) = \|\bm r\|_1+\min_{\sum_{j=1}^{\nu}k_j=\tau}W_{k_1k_2\cdots k_{\nu}}(\bm h,\bm r,{\bm a},{\bm d}).
\end{multline*} For similar reasons, if we let the cardinality of~$\mathcal{Y}$ change from~$0$ to~$n$, then we show that~$\min w(\mathcal{Y})-\|\bm r\|_1$ is equal to the minimum element of~$W(\bm h,\bm r,{\bm a},{\bm d})$. As a result, we conclude that the exponential number of inequalities described in~(\ref{formulacxy}) hold if and only if~$W(\bm h,\bm r,{\bm a},{\bm d})\geq O$. To summarize, the class~$\mathcal{A}(\bm h,\bm r, {\bm a},{\bm d})$ is nonempty if and only if~$W(\bm h,\bm r,{\bm a},{\bm d})\geq O$. This completes the proof.
\end{pf}

Theorem~\ref{tensorthm} gives rise to an intriguing physical interpretation of the supply/demand matching. Let us explain more with the simplest situation, when $\nu=1$ and the tensor condition reduces to~(\ref{OneOrderTensor}). Assume that~$k$ time slots have passed. For the supply side in the worst case, the first~$k$ time slots correspond to the largest~$k$ numbers in the supply profile, so the total remaining supply is given by the minuend of~(\ref{OneOrderTensor}), called the~$(k+1)$th supply tail. On the other hand, we consider the demand side in the best case, where each load can be charged at each of the first~$k$ time slots, and the remaining duration of load~$i$ becomes~$[r_i-k]^+$. Furthermore, the subtrahend of~(\ref{OneOrderTensor}) signifies the total remaining duration required by the loads, which is called the~\mbox{$(k+1)$th} demand tail. Clearly, if the supply is adequate, the~\mbox{$(k+1)$th} supply tail exceeds the~\mbox{$(k+1)$th} demand tail for all~\mbox{$0\leq k < n$}, while Theorem~\ref{tensorthm} states that such supply/demand tail dominance relationship is also sufficient to verify the adequacy of the supply. For the cases where~$\nu>1$, we observe similar interpretations. Rather than a scalar~$k$, each supply/demand tail pair is indexed by a vector of dimension~$\nu$, i.e.,~$[k_1~k_2~\cdots~k_\nu]$, where~\mbox{$0\leq k_\kappa\leq n_\kappa-n_{\kappa-1}$}, for each $\kappa\in \underline{\nu}$. Each element of the associated tensor is the difference between a certain supply/demand tail pair. Likewise, Theorem~\ref{tensorthm} shows that the uniform dominance relationship of all the indexed supply/demand tails implies the adequacy of the supply for the demand and vice versa.

\section{A Forward Market Implementation}
Economically, are the MAMD differentiated energy services practicable? With the procedures in~\cite{nayyar2016duration}, we explore such services in a forward market, where all the contracts are signed before the actual power delivery. We consider a continuum of consumers so as to avoid bin packing problems in allocation and assumptions on the concavity or uniformity of utility functions of different consumers, as suggested in~\cite{nayyar2016duration}. The three main elements involved are described as follows:
\begin{enumerate}[1)]
  \item Supply: In advance of transactions, the supplier knows the supply profile~$\bm{h}=[h_1~h_2~\cdots~h_n]$.
  \item Services: In this market, only the MAMD differentiated energy services are available and specified by~$\mathcal{S}$.
  \item Consumers: A continuum of loads are indexed by the points of a unit interval, namely, $x\in[0,1]$. In this case, each consumer~$x$ demands~$l(x)$ units/slot of power for~$r(x)$ time slots between the~\mbox{$(n_{a(x)}+1)$th} and the~$n_{d(x)}$th time slot, where~\mbox{$r(x),a(x),d(x)\in \mathbb{N}$} and~\mbox{$l(x)\!\in\mathbb{R}^+$ can be treated as functions over~$[0,1]$}. We can interpret~\mbox{$l(x)$ as} per capita demand, so the consumers in~\mbox{$[x,x+dx]$} demand~$l(x)dx$ units/slot of power for~$r(x)$ time slots.
      As a result, we shall denote the continuum of demand by~$$\mathcal{R}^c=\left\{\left(l(x),r(x),{a(x)},{d(x)}\right), x\in [0,1]\right\}.$$ Furthermore, the utility function of each consumer~$x$ in~$[0,1]$ is denoted by~$U\left(x,l(x),r(x),{a(x)},{d(x)}\right)$, where~$U\left(x,0,r(x),{a(x)},{d(x)}\right)=0$.
\end{enumerate}
We slightly generalize the structure tensor~$W(\bm h,\bm r,{\bm a},{\bm d})$ for a collection of discrete loads, and obtain the following structure tensor~$W^c(\bm h, \mathcal{R}^c)$ for the continuum of loads:
\begin{multline*}
  W^c_{k_1k_2\cdots k_{\nu}}(\bm{h},\mathcal{R}^c) =  \sum_{\kappa=1}^{\nu}\sum_{j=n_{\kappa-1}+k_\kappa+1}^{n_\kappa}h_j\\
 \ \ \ \ \ -\int_{0}^{1}\!\!\textcolor{mblue}{l(x)}\!\left[r(x)-k_{a(x)+1}-k_{a(x)+2}-\cdots-k_{d(x)}\!\right]^+dx,
\end{multline*}
where~$0\leq k_\kappa\leq n_\kappa-n_{\kappa-1}$, for all $\kappa\in \underline{\nu}$. Then, we state the next theorem which follows directly from Theorem~\ref{tensorthm}.
\begin{thm}\label{swmsol}
  The supply profile~$\bm h$ is adequate for the continuum of demand~$\mathcal{R}^c$ if and only if~$W^c(\bm h, \mathcal{R}^c)\geq O$.
\end{thm}

\subsection{Social Welfare Maximization} \label{socialwelfare}\vspace{-1pt}
Assuming there is a social planner who makes decisions for all involved parties, we wonder what the overall benefit is in this market. We define \emph{social welfare} as the summation of \emph{consumer welfare} and \emph{supplier revenue}. The consumer welfare is the difference between the consumers' total utilities and their expenditure for purchasing the MAMD services, while the supplier revenue is the difference between the gross profit from selling the MAMD services and the capitalized generation cost. As in~\cite{nayyar2016duration} and~\cite{negrete2016rate}, we assume the supplies are free. This is reasonable when the electrical energy is generated from renewable resources since the generation cost, mainly from infrastructure construction, is insensitive to the generation capacity and thus can be approximately regarded as a constant. As a result, the social welfare maximization problem is mathematically formulated as
\begin{equation}  \label{swm}
\begin{split}
\hspace{-4pt}\max_{l(\!x\!),r(\!x\!),{a(\!x\!)},{d(\!x\!)},x\in[0,1]}\!\!\!\!\!& \quad \int_{0}^{1}\!\!U\!\left(x,l(x),r(x),{a(x)},{d(x)}\right)\!dx\\
  \text{subject to~~~}  &\quad W^c(\bm h, \mathcal{R}^c)\!\geq\! O;~l(x)\geq 0, \\
  &\quad \left(r(x),{a(x)},{d(x)}\right)\!\in\! \mathcal{S}, \forall x\!\in\! [0,1].
  \end{split}
\end{equation}
Next, we will show that Problem~(\ref{swm}) has an optimal solution for any type of measurable utility function.

For each~$x\in [0,1]$, we define~$Z(x)=[Z_{k_1k_2\cdots k_{\nu}}(x)]$ with~$Z(0)=0$ and~$\dot{Z}(x)=[\omega_{k_1k_2\cdots k_{\nu}}(x)]$, where
\begin{equation}\label{diffw}
  \omega_{k_1k_2\cdots k_{\nu}}(x) = l(x)\!\left[r(x)-k_{a(x)+1}-\cdots-k_{d(x)}\!\right]^+\!,
\end{equation}
for all~$0\leq k_\kappa\leq n_\kappa-n_{\kappa-1}$, $\kappa\in \underline{\nu}$, and
\begin{equation}\label{diffw1}
  \omega_{k_1k_2\cdots k_{\nu}}(x) = U\left(x,l(x),r(x),{a(x)},{d(x)}\right),
\end{equation}
for all~$k_1= -1,0\leq k_\kappa\leq n_\kappa-n_{\kappa-1}$, $2\leq \kappa\leq \nu$.

Moreover, for~$x\in[0,1]$, we define the continuum of sets
 \begin{multline}\label{diff}
     \Omega(x)\!=\!\left\{\omega(x)\;|\; l(x)\geq 0,\left(r(x),{a(x)},{d(x)}\right)\!\in\! \mathcal{S}\right\}.
 \end{multline}
 Thus, we see a set-valued correspondence:~$x\mapsto\Omega(x)$ and~$\dot{Z}(x)\in\Omega(x), x\in [0,1].$
 By the integration of set-valued functions~\cite{aumann1965integrals} and combining~(\ref{diffw}),~(\ref{diffw1}) with (\ref{diff}), we denote the integral of the set-valued correspondence by~
 \begin{multline*}
     G= \int_0^1 \Omega(x)dx=\left\{Z(1)\;|\;Z(1) \text{ reached by a service}\right. \\
     \left.\text{allocation}~x\mapsto\left(l(x),r(x),{a(x)},{d(x)}\right), \forall x\in [0,1]\right\}.
 \end{multline*}
 With the first four theorems in~\cite{aumann1965integrals} on the integration of set-valued functions, we see that~$G$ is convex and closed.

 Then, we can redescribe the social welfare maximization problem with the notation regarding $Z(x)$ and $G$ so as to see whether the feasible region is compact: \begin{subequations}
\label{ReSocial}
 \begin{gather}
  \ \ \ \ \ \ \ \ \ \ \; \max_{Z(1)}\ \ \ \ \ \ \ \  Z_{(-1)0\cdots0}(1) \nonumber\\
  \text{subject to} ~~~~~~~~ Z(1)\in G;\label{inG}\\
 \label{zstruc}\begin{align}
  & Z_{k_1k_2\cdots k_\nu}(1)\leq  \sum_{\kappa=1}^{\nu}\sum_{j=n_{\kappa-1}+k_\kappa+1}^{n_\kappa}h_j,\nonumber \\
 &\ \ \ \text{ for all }0\leq k_\kappa\leq n_\kappa-n_{\kappa-1}, \kappa\in \underline{\nu}.
 \end{align}
 \end{gather}
 \end{subequations}
 Constraint~(\ref{zstruc}) consists of linear inequalities only. In light of this and the analysis of~$G$, the optimization variable $Z(1)$ is restricted to a compact and convex set by the constraints~(\ref{inG}) and~(\ref{zstruc}). It follows that an optimal solution to Problem~(\ref{ReSocial}) exists and thus so does the social welfare maximization problem~(\ref{swm}). Next, we will see whether the optimal social welfare of Problem~(\ref{swm}) can be obtained when the self-interested supplier and consumers separately maximize their own benefits.

\subsection{Competitive Equilibrium}\label{competitiveequilibrium}
The analysis herein is established in a perfectly competitive market, wherein every participant (the supplier and consumers) behaves as a rational and selfish price-taker. The price of a portion of the MAMD service~$(r,a,d)$ is $\pi_r^{a,d}$ and $\pi_0^{a,d}=0$. Facing a menu of services with the prices~$\left\{\pi_r^{a,d}\mid (r,a,d)\in \mathcal{S}\right\},$ each consumer orders the MAMD service and specifies the quantity, while the supplier decides whether an order is accepted, allocates the electrical energy over time slots, and adjusts the prices of MAMD services according to the supply and demand. The market dynamics are outside the scope of this paper. After the long-run evolution of this competitive market, the prices of MAMD services converge to an equilibrium, which should bring no incentives for rational participants to adjust their strategies~\cite{walras2013elements}.

For a market with the MAMD differentiated energy services described in this work, a competitive equilibrium is defined as a state satisfying the three conditions below:\vspace{-2pt}
\begin{enumerate}[1)]
\item Each consumer maximizes its welfare. Precisely, for the continuum of loads in~$[0,1]$, each consumer~$x$ selects~the parameters~$l(x)$,~$r(x)$,~${a(x)}$, and~${d(x)}$ to maximize its net benefit. This leads to a continuum of welfare maximization problems: for all~$x\in[0,1]$,
     \begin{multline}\label{cpdemand}
 \max_{l(x)\geq 0,(r(x),{a(x)},{d(x)})\in \mathcal{S}}\Big(U\left(x,l(x),r(x),{a(x)},{d(x)}\right)\\-l(x)\pi_{r(x)}^{{a(x)},{d(x)}}\Big).
      \end{multline}
\item The supplier maximizes its revenue. To be specific, given the MAMD service prices~$\left\{\pi_r^{a,d}\mid (r,a,d)\in \mathcal{S}\right\}$ and constrained by the available supply~$\bm h$, the supplier decides the quantity~$q_r^{a,d}$ of each MAMD service~$(r,a,d)\in \mathcal{S}$ to be produced with the purpose of maximizing its revenue. Mathematically, defining the service time interval set:
     \begin{equation*}
      \mathcal{F}=\left\{(a,d)~|~a,d\in\mathbb{N},  a< d\leq \nu\right\},
     \end{equation*} we formulate the revenue maximization problem as
     \begin{align}\label{cpsupply}
    \max_{q_r^{a,d}}& \quad\sum_{(a,d)\in \mathcal{F}}\sum_{r=1}^{n_d-n_a}q_r^{a,d}\pi_r^{a,d} \nonumber\\
    \mbox{subject to}
    &\quad \quad\delta_j^{a,d} = \sum_{r=j}^{n_d-n_a}q_r^{a,d}; \\
    &\hspace{-60pt} \sum_{\kappa=1}^{\nu}\sum_{j=n_{\kappa-1}+k_\kappa+1}^{n_\kappa}\!\!\!\!h_j-\!\!\!\!\sum_{(a,d)\in\mathcal{F}}\sum_{j=k_{a+1}+\cdots+k_{d}+1}^{n_d-n_a}\!\!\!\!\!\delta_j^{a,d}\geq 0,\nonumber\\
    &\text{for all } 0\leq k_\kappa\leq n_\kappa-n_{\kappa-1}, \kappa\in \underline{\nu}.\nonumber
      \end{align}
     In particular, the above constraints address the adequacy condition, following from Theorem~\ref{swmsol}.

\item The market is clear. This means that the supply and demand balance out, namely, for each~$(r,a,d)\in \mathcal{S}$,
  \begin{equation}\label{cpbalance}
    \hspace{-29pt}q_r^{a,d}\!=\!\!\int_0^1\!\!l(x)\mathds{1}(r(x)\!=\!r,{a(x)}\!=\!a,{d(x)}\!=\!d)dx.\!\!\!\!\!\!\!\!\!\!\!
  \end{equation}
\end{enumerate}\vspace{-6pt}

The analysis regarding the competitive equilibrium relies on the assumption that an individual transaction does not influence the prices. First of all, we wonder whether a competitive equilibrium exists in a forward market with MAMD services only.
\begin{thm}\label{efficiency}
  There exists a menu of MAMD service prices~$\left\{\pi_r^{a,d}\mid (r,a,d)\in \mathcal{S}\right\}$, such that Problem~(\ref{cpdemand}) and Problem~(\ref{cpsupply}) admit optimal solutions satisfying~(\ref{cpbalance}).
\end{thm}
\vspace{-8pt}
\begin{pf}
Denote the optimal solution to Problem~(\ref{swm}) by the service allocation~$x\mapsto\left(\bar{l}(x),\bar{r}(x),{\bar{a}(x)},{\bar{d}(x)}\right)$, for all~$x\in [0,1]$ and that to Problem~(\ref{ReSocial}) by~$\bar{Z}(1)$. The overbar indicates realizations of related symbols. Dualize Problem~(\ref{ReSocial}) regarding the constraints described by (\ref{zstruc}). As a result, there exist a bundle of Lagrange multipliers: $$\alpha_{k_1k_2\cdots k_{\nu}}\!\geq 0\text{, for all } 0\leq k_\kappa\leq n_\kappa-n_{\kappa-1}, \kappa\in \underline{\nu},$$ such that $\bar{Z}(1)$ is also the optimal solution to the following optimization problem:
  \begin{align}\label{dual}
    \max_{Z(1)\in G} Z_{-1\ 0\  \cdots\  0}(1)-\!\!\!\!\!\!\!\sum_{k_1,k_2,\dots,k_{\nu}}\!\!\!\!\!\!\alpha_{k_1k_2\cdots k_{\nu}}Z_{k_1k_2\cdots k_{\nu}}(1).
  \end{align}
  In addition, it follows from the complementary slackness that, for all~$0\leq k_\kappa\leq n_\kappa-n_{\kappa-1}, \kappa\in \underline{\nu}$,
  \begin{multline}\label{slack}
    \hspace{-5pt}\alpha_{k_1k_2\cdots k_{\nu}}\!\!\left[\!\bar{Z}_{k_1k_2\cdots k_{\nu}}(1)- \sum_{\kappa=1}^{\nu}\sum_{j=n_{\kappa-1}+k_\kappa+1}^{n_\kappa}\!\!h_j\!\right]\!=\!0.
  \end{multline}
  According to~(\ref{diffw}) and~(\ref{diffw1}), we rewrite the term to be maximized in (\ref{dual}) and obtain
  \begin{equation}\label{remax}
    \int_0^1\!\!U\!\left(x,l(x),r(x),{a(x)},{d(x)}\right)\!-l(x)\bar{\pi}_{r(x)}^{{a(x)},{d(x)}}dx,
  \end{equation}
 \begin{equation} \text{where}~\bar{\pi}_{r(x)}^{{a(x)},{d(x)}}\!=\!\!\!\!\!\!\!\!\sum_{k_1,k_2,\dots,k_{\nu}}\!\!\!\!\!\!\alpha_{k_1k_2\cdots k_{\nu}}\!\!\left[\!r(x)\!-\!\!\!\!\!\!\!\sum_{\kappa=a(x)+1}^{d(x)}\!\!\!k_\kappa\!\right]^+\!\!.\label{price}
  \end{equation}
  In what follows, we shall explain why we above use the notation~$\bar{\pi}$ related to the MAMD service prices. Firstly, we consider the consumer welfare maximization. Treat the term~(\ref{remax}) as the summation of an infinite number of sub-terms indexed by~$x\in[0,1]$, and all these sub-terms are independent of each other. From this point of view, since~$\bar{Z}(1)$ solves~(\ref{dual}), it is not difficult to see that
  \begin{multline*}
    \left(\bar{l}(x),\bar{r}(x),{\bar{a}(x)},{\bar{d}(x)}\right)\!=\\ \arg \!\max_{l(x)\geq 0,(r(x),{a(x)},{d(x)})\in \mathcal{S}}\Big\{U\left(x,l(x),r(x),{a(x)},{d(x)}\right)\\-l(x)\bar{\pi}_{r(x)}^{{a(x)},{d(x)}}\Big\},
  \end{multline*}
  where~$\bar{\pi}_{r(x)}^{{a(x)},{d(x)}}$ is calculated by~(\ref{price}). As a consequence, no consumer will intend to violate the service allocation~$x\mapsto\left(\bar{l}(x),\bar{r}(x),{\bar{a}(x)},{\bar{d}(x)}\right),~x\in[0,1]$ under the prices given by~(\ref{price}). That is, the continuum of welfare maximization problems described in~(\ref{cpdemand}) are simultaneously solved by such a service allocation.

  Then, for a clear market, under the mentioned service allocation~$x\mapsto\left(\bar{l}(x),\bar{r}(x),{\bar{a}(x)},{\bar{d}(x)}\right)\text{, for all}~x\in[0,1]$, the supplier should produce a bundle of MAMD services described by~$\left\{\bar{q}_r^{a,d}\mid (r,a,d)\in \mathcal{S}\right\}$, where~\begin{equation}\bar{q}_r^{a,d}=\int_0^1 l(x)\mathds{1}(\bar{r}(x)=r,{\bar{a}(x)}\!=a,{\bar{d}(x)}=d)dx.\label{condition3}\end{equation}
  Finally, to complete the proof, it remains to show that under the prices calculated by Formula~(\ref{price}), the above bundle of MAMD differentiated energy services not only maximize the supplier revenue, but can also be generated from the given supply~$\bm h$.

  For this purpose, we restate the supplier revenue as
  \begin{equation*}
    \sum_{(a,d)\in\mathcal{F}}\!\sum_{r=1}^{n_d-n_a}\bar{q}_r^{a,d}\bar{\pi}_r^{a,d}=\!\!\sum_{(a,d)\in\mathcal{F}}\!\sum_{j=1}^{n_d-n_a}\bar{\delta}_j^{a,d}\Big(\bar{\pi}_j^{a,d}-\bar{\pi}_{j-1}^{a,d}\Big),
  \end{equation*}
  where $\bar{\delta}_j^{a,d}=\sum_{j=r}^{n_d-n_a}\bar{q}_r^{a,d}$. Substituting the expression of the price (\ref{price}) into the above formula yields
  \begin{equation*}
    \sum_{(a,d)\in\mathcal{F}}\sum_{r=1}^{n_d-n_a}\bar{q}_r^{a,d}\bar{\pi}_r^{a,d}
  \end{equation*}
  \noindent \begin{align*}
    &= \sum_{k_1,k_2,\dots,k_{\nu}}\!\!\!\!\!\alpha_{k_1k_2\cdots k_{\nu}}\!\left(\!\sum_{(a,d)\in\mathcal{F}}\sum_{j=k_{a+1}+k_{a+2}+\cdots+k_d+1}^{n_d-n_a}\!\!\!\bar{\delta}_j^{a,d}\!\!\right)\\
    &\leq \sum_{k_1,k_2,\dots,k_{\nu}}\!\!\!\!\!\alpha_{k_1k_2\cdots k_{\nu}}
    \left(\sum_{\kappa=1}^{\nu}\sum_{j=n_{\kappa-1}+k_\kappa+1}^{n_\kappa}h_j\right)\\
    &= \sum_{k_1,k_2,\dots,k_{\nu}}\!\!\!\!\!\alpha_{k_1k_2\cdots k_{\nu}}\bar{Z}_{k_1k_2\cdots k_{\nu}}(1).
  \end{align*}
  The inequality above follows from the adequacy constraint which is described by the nonnegativity of an associated structure tensor, while the last equality follows from~(\ref{slack}) by the complementary slackness. To this point, we have shown that the production bundle~$\left\{\bar{q}_r^{a,d}\mid (r,a,d)\in \mathcal{S}\right\}$ can be generated by the given supply~$\bm h$ and attains the maximum supplier revenue.

  To summarize, we see that under the menu of MAMD service prices~$\left\{\bar{\pi}_r^{a,d}\mid (r,a,d)\in \mathcal{S}\right\}$, the service allocation~$x\mapsto\left(\bar{l}(x),\bar{r}(x),{\bar{a}(x)},{\bar{d}(x)}\right)$, for all~$x\in [0,1]$ solves a continuum of problems described by~(\ref{cpdemand}), while the bundle of MAMD services~$\left\{\bar{q}_r^{a,d}\mid (r,a,d)\in \mathcal{S}\right\}$ solves Problem~(\ref{cpsupply}). Finally, we complete the proof by~(\ref{condition3}).
\end{pf}
In economic analysis, we usually regard a competitive equilibrium as a measure of the market efficiency~\cite{arrow1954existence}. Specifically, a competitive equilibrium is efficient if the social welfare attained in a distributed manner is equal to the optimal objective of the social welfare maximization problem~(\ref{swm}). From the proof of Theorem~\ref{efficiency}, we have proven that the optimal solution to Problem~(\ref{ReSocial}) helps generate a menu of equilibrium prices, which result in a state satisfying the three conditions for a competitive equilibrium. Consequently, we also establish the existence of an efficient competitive equilibrium. This verifies the economic feasibility of MAMD services.

In above proof, we apply the Lagrange multiplier method, which gives a heuristic economic interpretation. By~(\ref{price}), the Lagrange multipliers work as implicit price factors. Specifically, each multiplier indexed by~$[k_1~k_2~\cdots~k_\nu]$ adds a weight to a term of demand tail,~$[r-\sum_{\kappa=a+1}^{d}k_\kappa]^+$. To sum up, the price of the MAMD service~$(r,a,d)$ is given by the summation of the weighted terms over~$[k_1~k_2~\cdots~k_\nu]$. From~(\ref{price}), we can get a number of practical insights which are consistent with our intuition that the less laxity the MAMD service has, the higher the price of the service will be. For example, \smallskip \newline
(1)~With a fixed~$(a,d)\in \mathcal{F}$, the price function $\pi_r^{a,d}$ is nondecreasing as the duration~$r$ increases.\\
(2)~Given~$\left({a},{b}\right),\left({c},{d}\right)\in \mathcal{F}$, if ${a}\!\geq\! {c}$ and ${b}\!\leq\! {d}$, then it follows that~$\pi_r^{{a},{b}}\geq \pi_r^{{c},{d}}$, for~$0<r \leq n_{b}-n_{a}.$

To a certain degree, the existence of an efficient competitive equilibrium allows us to skirt the debate on a planned economy versus a market economy. As shown in~\cite{chen2015duration},~\cite{nayyar2016duration}, and this paper, it is an accepted practice to check whether a new market is well-defined theoretically by exploring an efficient competitive equilibrium. In an efficient market, each participant makes decisions for its own benefit, but the market dynamics can converge to a socially optimal status automatically. Although these results rest on a perfectly competitive price system in a forward market, they reveal the potential for the success of MAMD services in a practical market implementation. {Moreover, numerically finding an efficient competitive equilibrium is challenging and thus left for future work. Nevertheless, we can construct the equilibrium contracts analytically for certain illustrative cases as shown in~\cite{nayyar2016duration}.}

\section{Simulation Results}
{We show by simulation that the MAMD services better ease the burden on the supplies than other conceivable benchmark models, by allowing different arrival times and deadlines.} We use synthesized data to avoid cumbersome technical details, while the techniques in~\cite{negrete2016rate} and~\cite{schmidt2018efficiently} can be used to fit the MAMD model to real data.

Let us consider a parking slot in a neighborhood. The operational horizon ranges from~$6$~p.m. to~$10$~a.m (next day). The specified arrival times and deadlines are~$6$~p.m.,~$9$~p.m.,~$1$~a.m.,~$6$~a.m.,~$8$~a.m., and~$10$~a.m. If each time slot accounts for one hour, then we can fit such a scenario to the MAMD model by setting parameters as~$\nu=5$,~$n_0=0$,~$n_1=3$,~$n_2=7$,~$n_3=12$,~$n_4=14$, and~$n_5=n=16$. As a result, the MAMD service~$(6,1,4)$ means that it can charge a load for six hours from~$9$~p.m. to~$8$~a.m. the next day. If there is a visitor who arrives before~$6$~p.m. and leaves after~$1$~a.m., she may require the MAMD service~$(3,0,2)$ to charge her car for three hours.

For contrast, we take the duration-differentiated energy services~\cite{nayyar2016duration} as the benchmark. For comparison with the duration-deadline jointly differentiated energy services, a similar analysis can be obtained and is thus omitted for brevity. If we consider a single benchmark model, then many loads served by MAMD services cannot be accommodated by the benchmark model. For example, if we let the duration-differentiated energy services start from~$6$~a.m. and end at~$10$~a.m., then the supplier has to reject the loads which must arrive after~$6$~a.m. or depart before~$10$~a.m. In this situation, it is easy to see that the MAMD model can accommodate more kinds of flexible loads when we only consider a single benchmark model.

For better comparison, we consider five benchmark models together. Their operational horizons are in order from~$6$~p.m. to~$9$~p.m.~\mbox{($n=3$)}, from~$9$~p.m. to~$1$~a.m.~\mbox{($n=4$)}, from~$1$~a.m. to~$6$~a.m.~\mbox{($n=5$)}, from~$6$~a.m. to~$8$~a.m.~\mbox{($n=2$)}, and from~$8$~a.m. to~$10$~a.m.~\mbox{($n=2$)}. As a result, loads requiring MAMD services can buy a combination of benchmark services from the five benchmark models instead. For example, a load requiring the MAMD service $(3,0,2)$ can require two charging hours from the first benchmark model and one charging hour from the second one. Because this load is indifferent of the actual power delivery time, it treats equally such combined benchmark services and the MAMD service~$(3,0,2)$.

We herein consider supply profiles which are adequate for load collections requiring MAMD services. Then, each load replaces its MAMD service with an equivalent combination of benchmark services. Such replacements may not be unique. For example, the MAMD service~$(3,0,2)$ is also identical to one charging hour in the first benchmark model and two hours in the second one. In this case, the load will randomly pick a replacement.

\begin{figure}[t]
\begin{minipage}[t]{0.49\linewidth}
   \centering
   \includegraphics[scale=0.315]{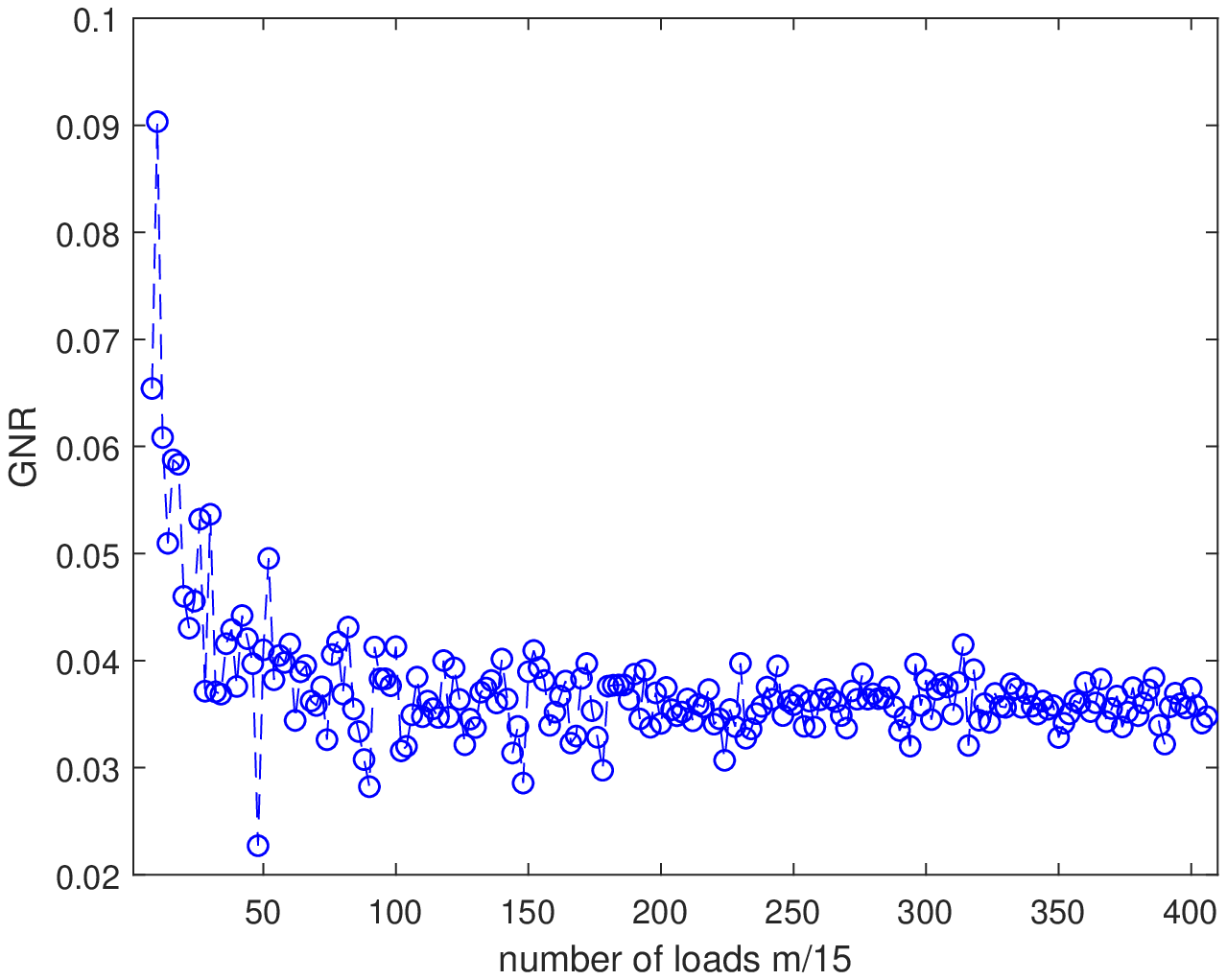} 
\end{minipage}
\begin{minipage}[t]{0.49\linewidth}
   \centering
   \includegraphics[scale=0.315]{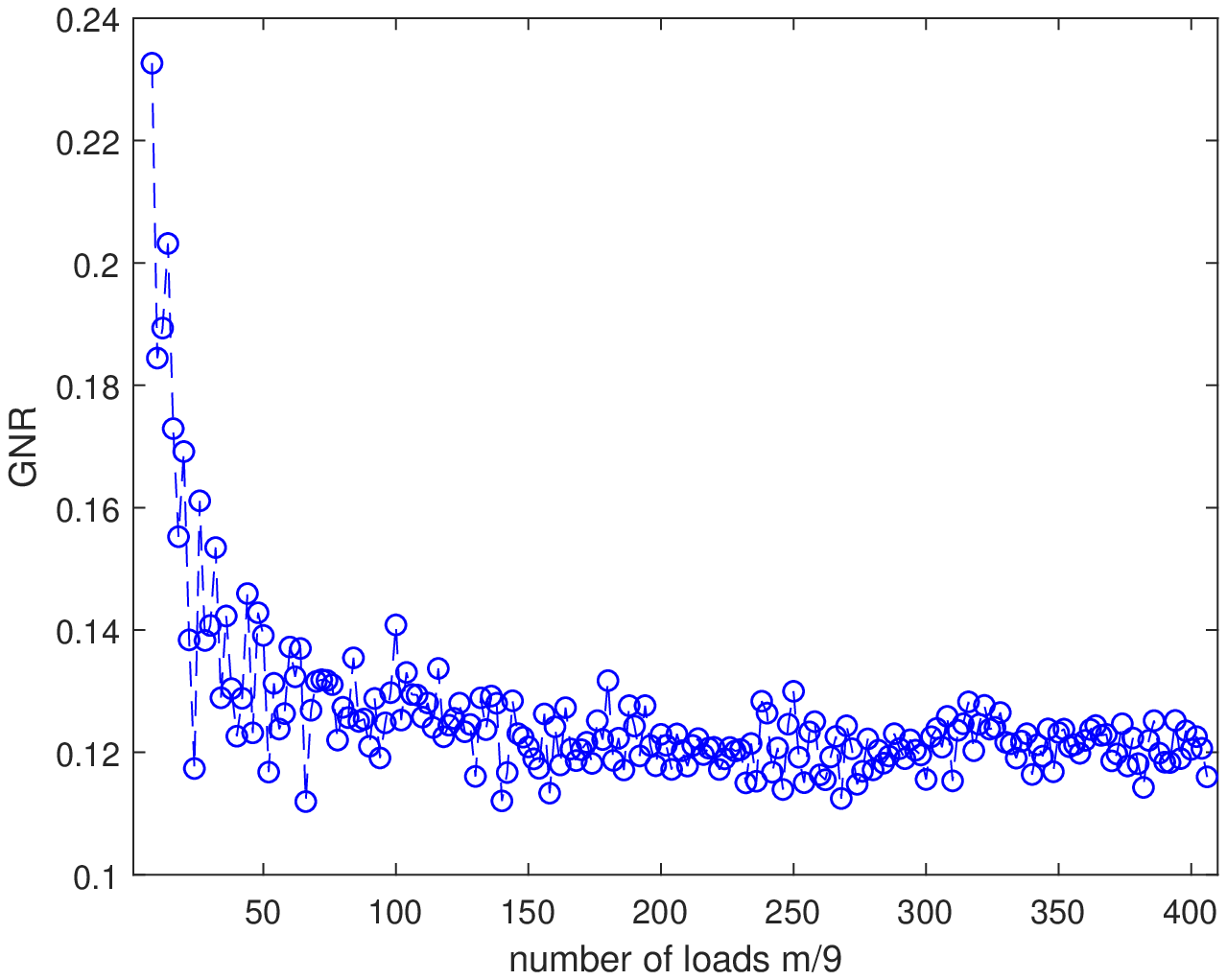} 
\end{minipage}
\caption{GNR under~fifteen kinds of MAMD services~(left) and GNR under~nine kinds of MAMD services~(right).}
   \label{fig: tworatio}
\end{figure}

The supply may become inadequate if we transfer the demand from MAMD services to combinations of benchmark services. {Thus, we consider the model-adequacy gap, which is the minimum amount of energy supplementary to the insufficient supplies of the benchmark models so that the augmented combined supply is adequate.} For a fixed arrival-deadline pair~$(a,d)\in \mathcal{F}$, we randomly generate the duration $r$. Theoretically, there are fifteen arrival-deadline pairs in total for the considered MAMD services. We generate an equal number of loads for each arrival-deadline pair. Let GNR denote the ratio of the model-adequacy gap to the number of loads. As depicted on the left of Figure~\ref{fig: tworatio}, the GNR approximately converges to~$3.5\%$. On the right of Figure~\ref{fig: tworatio}, we merely consider nine kinds of MAMD services in terms of the arrival-deadline pair, and the GNR still approximately converges to~$12\%$. Due to the convergence, we conclude that the service provider will suffer more losses by transferring the MAMD services to the benchmark services, as the number of loads increases.

\section{Conclusions and Future Work}
In this paper, we investigate the market implementation of MAMD differentiated energy services. We theoretically confirm the economic feasibility of such services by proving the existence of an efficient competitive equilibrium in a forward market. That is, the distributive solution in the competitive price system is consistent with the centralized one attained by a social planner.

In the future, we will conduct a data-driven analysis of differentiated energy services, which involves more practical concerns, e.g., how to obtain credible data on the demand. Regarding the economic analysis, we will pay more attention to the market design together with the market dynamics. Moreover, many technical issues remain to be discussed regarding optimal energy coordination for flexible loads~\cite{mo2017coordinating} and how to apply differentiated energy services more efficiently in the presence of uncertainties on the supply and demand~\cite{mo2018staircase}.

\begin{ack}                               
The authors would like to thank Dr.~Sei Zhen Khong of The University of Hong Kong and Dr. Hao Wang of Stanford University for helpful discussions. 
\end{ack}

\bibliographystyle{ieeetr}          
\bibliography{DESGMB}           

\end{document}